# A Novel Approach to Document Classification using WordNet

**Koushiki Sarkar[#] and Ritwika Law[*]**

**ABSTRACT**

Content based Document Classification is one of the biggest challenges in the context of free text mining. Current algorithms on document classifications mostly rely on cluster analysis based on bag-of-words approach. However that method is still being applied to many modern scientific dilemmas. It has established a strong presence in fields like economics and social science to merit serious attention from the researchers. In this paper we would like to propose and explore an alternative grounded more securely on the dictionary classification and correlatedness of words and phrases. It is expected that application of our existing knowledge about the underlying classification structure may lead to improvement of the classifier's performance.

## 1. INTRODUCTION

Content based Document Classification is one of the biggest challenges in the context of free text mining. This is a problem relevant to many areas of Physical and Social Sciences. At a basic level, it is a challenge of identifying features, useful for classification, in qualitative data. At a more applied level, it can be used for classifying sources (Human, Machine or Nature) of such data. Among the more important recent applications, we note its usage in the Social Networking sites (see [1], [2], [5] and [6]), Medical Sciences [3] and Media [4] among others.

One of the major problems with most data classification models is that the classification is essentially blind. Current algorithms on document classifications mostly rely on cluster analysis based on bag-of-words approach. The basic classifiers that use a bag-of-words approach, or more sophisticated Bayesian classification algorithms all mostly use word frequency in one form or the word. Word sense is almost always ignored. Such methods rely on an ad-hoc idea of the correlatedness between words. While the blind approach should work if we have a documents of reasonable size or a large corpus to begin with, so that we have an easier time picking up signatures of "good" or "bad" sets, such a method may not work with a smaller size of the set. An author may use a diverse vocabulary, so that overall frequency of good or bad words are low, thereby making classification harder. For example, if "good" is a positive word and in the document the author uses the term "awesome " multiple times and "good" never, we may not capture the document author's positive sentiment by our mechanism. This becomes an especially relevant concern while sentiment extraction from smaller documents, like a tweet or a product review at an online sight.

However such methods are still being applied to many modern scientific dilemmas because of the urgency of the problems. It has established a strong presence in fields like economics and social science to merit serious attention from the researchers. In this paper we would like to propose and explore an alternative grounded more securely on the dictionary classification and correlatedness of words and phrases. It is expected that application of our existing knowledge about the underlying classification structure may lead to improvement of the classifier's performance.

---

[#] Indian Statistical Institute, Kolkata, koushikisarkars@gmail.com
[*] Calcutta University, Kolkata, ritwika.ambika.law2@gmail.com

**1.1 Abstraction of Document in Classification Problem:**

In a typical problem, we are given a set of documents and two prefixed classes in which the documents have to be classified. We have to develop an optimum rule for this classification. The collection of documents is called a corpus.

There are multiple ways to visualize a document, the most common among which is the bag-of-words approach. Let our corpus be called c, which contains n documents, namely $\{d_1, d_2, ..., d_n\}$. Each $d_i$ contains a finite collection of words, call them $\{w_{i1}, w_{i2}, ..., w_{in}\}$.

In Bag-of-Words approach, we can consider the words in conjunction with their frequencies in the document, which, after stemming, are used for classification. However, this destroys the ordering of words which may lead to higher misclassification probability. It makes more sense to consider the document as a finite sequence of words where repetitions are possible.

Our approach is as follows. Given a training dataset, and assuming a binary classification setup into categories A and B representing good and bad respectively; from our training dataset we can construct two distinct weighted networks of words and phrases that represent the "closeness" of the words and eventually helps us to decide the classification of a document based on the closeness of the contents of this new document with either the "good" network or the "bad". Thus, in the end, each document belongs to one of the two possible classes - category A or category B. We also extract a set of correlated words or phrases as our features set to be used for further classification.

Here we will be using WordNet which provides a semantic lexicon for English. WordNet is a large lexical database of English. Nouns, verbs, adjectives and adverbs are grouped into sets of cognitive synonyms (synsets), each expressing a distinct concept. Synsets are interlinked by means of conceptual-semantic and lexical relations. The resulting network of meaningfully related words and concepts can be navigated with the browser. WordNet's structure makes it a useful tool for computational linguistics and natural language processing.

**2. *Structure of WordNet***

WordNet is a lexical dictionary available online free of cost. WordNet is somewhat of an extended version of a thesaurus. The main relation among words in WordNet is synonymy, as between the words *shut* and *close* or *car* and *automobile*. Synonyms--words that denote the same concept and are interchangeable in many contexts--are grouped into unordered sets (synsets). Each of WordNet's 117 000 synsets is linked to other synsets by means of a small number of "conceptual relations." Additionally, a synset contains a brief definition and, in most cases, one or more short sentences illustrating the use of the synset members. Word forms with several distinct meanings are represented in as many distinct synsets. Thus, each form-meaning pair in WordNet is unique.

Both nouns and verbs are organized into hierarchies, defined by [hypernym](hypernym) or *[IS A](IS A)* relationships. For instance, one sense of the word *dog* is found following hypernym hierarchy; the words at the same level represent synset members. Each set of synonyms has a unique index.

dog, domestic dog, Canisfamiliaris

```
  =>canine, canid
   =>carnivore
     =>placental, placental mammal, eutherian, eutherian mammal
      =>mammal
       =>vertebrate, craniate
        =>chordate
         =>animal, animate being, beast, brute, creature, fauna
          => ...
```

At the top level, these hierarchies are organized into 25 beginner "trees" for nouns and 15 for verbs (called *lexicographic files* at a maintenance level). All are linked to a unique beginner synset, "entity." Noun hierarchies are far deeper than verb hierarchies

Adjectives are not organized into hierarchical trees. Instead, two "central" antonyms such as "hot" and "cold" form binary poles, while 'satellite' synonyms such as "steaming" and "chilly" connect to their respective poles via a "similarity" relations. The adjectives can be visualized in this way as "dumbbells" rather than as "trees."

## 2.1 Notion of Semantic Similarity

It is easy to define a similarity measure between word pairs via WordNet. WordNet already has a few existing Perl modules. Wordnet:similarity uses a "is-as" relationship between nouns to classify them in the same synset. For example, "dog" and "animal" are closer than "dog" and "closet" is. Also another point to note is that this "is-as" relationship does not cross parts of speech boundary.

This, however, only captures a small notion of similarity between words as there can be many other relations aside from "is-as". WordNet also contains other non-hierarchical relationships between words which are expanded upon in a "gloss" or definition added.

### 2.1.1 Path Similarity

There are multiple notions of similarity possible in WordNet Lexicography. The three major ones, based on path length, are:

- lch (Leacock & Chodorow 1998) measure = $\frac{d(a,b)}{\max d(a,b)}$ where $d(.,.)$ is the shortest path between two concepts a and b in the "is-a" system.
- wup (Wu & Palmer 1994) measure = $\frac{d(\text{root node, LCS})}{d(\text{root node, a})+d(\text{root node, b})}$ where LCS or least common subsumer of the two concepts is the most specific concept they have as an ancestor.
- path measure = $\frac{1}{d(a,b)}$, i.e., the path measure is equal to the inverse of the shortest path length between two concepts.

In this regard we start to view documents as a point in the co-ordinate system where X-axis indicates the degree of inclination towards the bad set and Y-axis indicates the degree of inclination towards the good set.

## 2.2 Operational Approach

A document can be considered as an ordered conglomeration of words. We start with n documents and with two existing groups of words to be used for classification. Call them $w_{0G} = \{w_{1G},...,w_{nG}\}$ and $w_{0B} = \{w_{1B},...,w_{nB}\}$.

This wordlist can be given to us, or be captured from a training set. Suppose the wordlist is given. We can then proceed for classification. Pick the document $d_i = \{w_{i1}, w_{i2},...,w_{in}\}$. For each word in the document we calculate the distances from the words of $w_{0G}$ and $w_{0B}$. We consider the proportion of classification of words to each group. We prefix $\{\epsilon_1, \epsilon_2\}$ in such a way that $p_A > \epsilon_1$ we classify to group 1; $p_B > \epsilon_2$ we classify to group 2, anything in between we fail to classify.

Each word from the WordNet are taken and its distance from categories A is compared with the distance from category B. The distance between two words is considered in terms of number of nodes (intermediate words) between them. If the distance from the category A is greater than the distance from category B then the word which has been taken from the WordNet will be a similar sounding word of category B and thus it will get appended to category B otherwise it gets appended to category A. Like this we can split the words into two categories and expand the given two set of words which will be used later to classify a data. The diagram below illustrates the idea of and leads to the Algorithm that we develop.

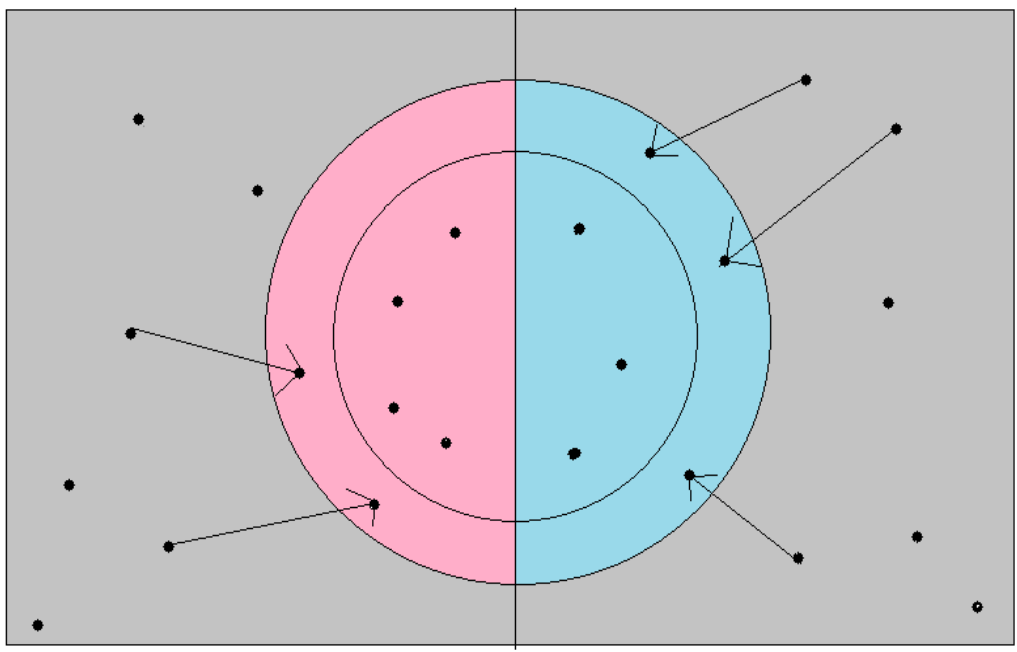

Figure 1: Inserting words into the two sets- good and bad from the WordNet. The pink colour represents good set, the blue colour represents bad set and the grey represents the words of the WordNet which are not present in either of the two sets. The words present in the grey portion and in the first half are similar sounding words of the set good. The distance of these words from the good set is lesser than that of bad set so these are appended to the good set. The words whose distance from the bad set is lesser than that of good set are placed at the second half of the grey portion. These are appended to the bad set. All the words are not appended. Only the words satisfying a tau condition check are appended. This is done to avoid unnecessary increase of the sets.

**2.3 Graphical Position of a Document in a Appropriate Coordinate System:**

It is not essential that every document can be sufficiently polarized to have a strong inclination towards to a particular word group. Also, the information content of one document may be higher than that of another document. Say, two newspapers may both support a political party, but one is subtle and another is more vocal. Simply classifying them to the same group ignores the distance in opinion among them. This may also lead to high degree of misclassification in sparse or small sized documents, like tweets, where our strength of evidence is low. In such a case it makes more sense to also report our degree of belief about the classification aside from the class itself. The strength of evidence can be calculated by many metrics, we use the proportion of words classified to group as the strength. Thus, depending on situation, a document may convey strong feelings in favour of both groups simultaneously- for example, an IMDB review that criticizes the action sequences in a movie yet praises the screenplay.

Starting from the selected list of "good" and "bad" words, it becomes imperative to create a classification method from both these wordlists. We need to find, in our test datasets, a frequency based measure for this classification. So, starting from the original set, we pick each document with the good words in them and the bad words in them and calculate the frequency of these words in each of the documents. A proportion of words classified as "good" and "bad" weighted by their frequencies is used for classification. A diagrammatic representation of the algorithm to capture the frequencies of the words is given below.

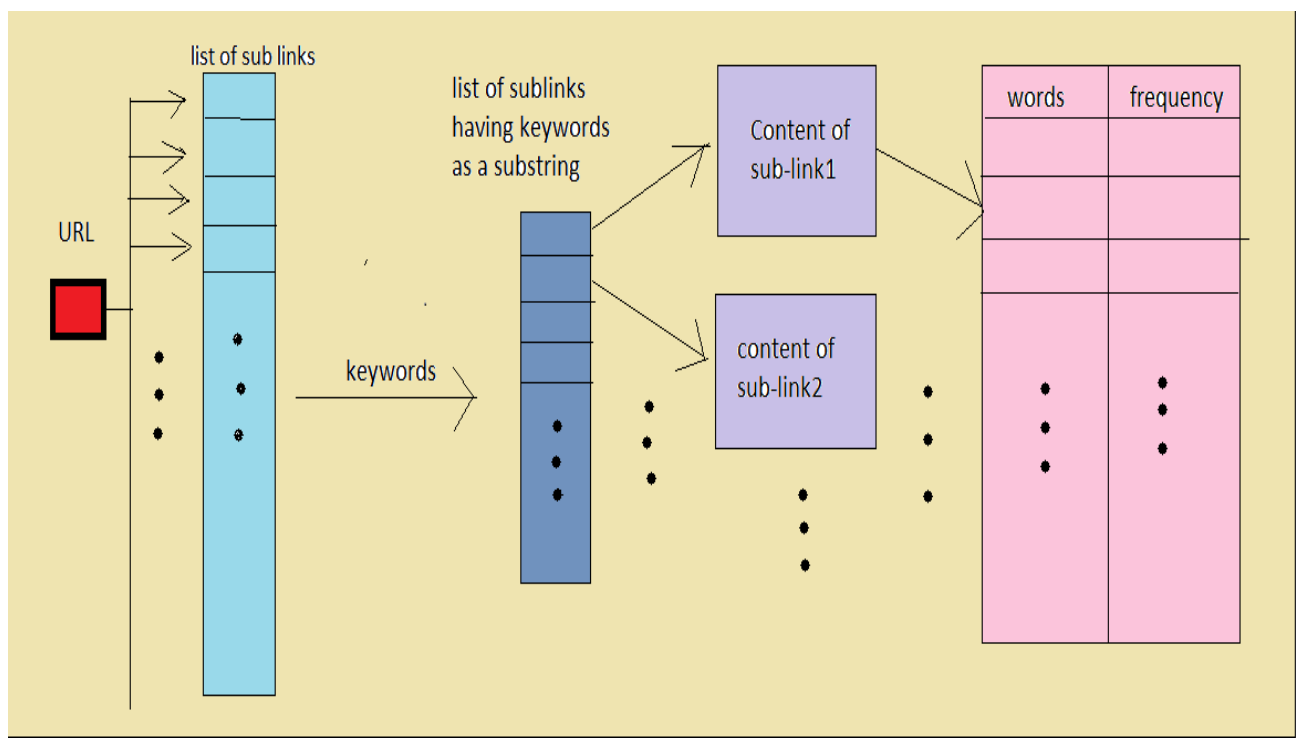

Figure 2: A Diagrammatic Representation of the Calculation of Word Frequency for the List of Words generated in Part 1.

### 2.4 Example of WordNet Search - 3.1
Word to search for: | good |

Display Options:
Key: "S:" = Show Synset (semantic) relations, "W:" = Show Word (lexical) relations
Display options for sense: (gloss) "an example sentence"

**Noun**

- S: (n) **good** (benefit) *"for your own good"; "what's the good of worrying?"*
- S: (n) **good**, goodness (moral excellence or admirableness) *"there is much good to be found in people"*
- S: (n) **good**, goodness (that which is pleasing or valuable or useful) *"weigh the good against the bad"; "among the highest goods of all are happiness and self-realization"*
- S: (n) commodity, trade good, **good** (articles of commerce)

**Adjective**

- S: (adj) **good** (having desirable or positive qualities especially those suitable for a thing specified) *"good news from the hospital"; "a good report card"; "when she was good she was very very good"; "a good knife is one good for cutting"; "this stump will make a good picnic table"; "a good check"; "a good joke"; "a good exterior paint"; "a good secretary"; "a good dress for the office"*
- S: (adj) full, **good** (having the normally expected amount) *"gives full measure"; "gives good measure"; "a good mile from here"*
- S: (adj) **good** (morally admirable)
- S: (adj) estimable, **good**, honorable, respectable (deserving of esteem and respect) *"all respectable companies give guarantees"; "ruined the family's good name"*
- S: (adj) beneficial, **good** (promoting or enhancing well-being) *"an arms limitation agreement beneficial to all countries"; "the beneficial effects of a temperate climate"; "the experience was good for her"*
- S: (adj) **good** (agreeable or pleasing) *"we all had a good time"; "good manners"*
- S: (adj) **good**, just, upright (of moral excellence) *"a genuinely good person"; "a just cause"; "an upright and respectable man"*
- S: (adj) adept, expert, **good**, practiced, proficient, skillful, skilful (having or showing knowledge and skill and aptitude) *"adept in handicrafts"; "an adept juggler"; "an expert job"; "a good mechanic"; "a practiced marksman"; "a proficient engineer"; "a lesser-known but no less skillful composer"; "the effect was achieved by skillful retouching"*
- S: (adj) **good** (thorough) *"had a good workout"; "gave the house a good cleaning"*
- S: (adj) dear, **good**, near (with or in a close or intimate relationship) *"a good friend"; "my sisters and brothers are near and dear"*
- S: (adj) dependable, **good**, safe, secure (financially safe) *"a good investment"; "a secure investment"*
- S: (adj) **good**, right, ripe (most suitable or right for a particular purpose) *"a good time to plant tomatoes"; "the right time to act"; "the time is ripe for great sociological changes"*
- S: (adj) **good**, well (resulting favorably) *"it's a good thing that I wasn't there"; "it is good that you stayed"; "it is well that no one saw you"; "all's well that ends well"*
- S: (adj) effective, **good**, in effect, in force (exerting force or influence) *"the law is effective immediately"; "a warranty good for two years"; "the law is already in effect (or in force)"*
- S: (adj) **good** (capable of pleasing) *"good looks"*
- S: (adj) **good**, serious (appealing to the mind) *"good music"; "a serious book"*
- S: (adj) **good**, sound (in excellent physical condition) *"good teeth"; "I still have one good leg"; "a sound mind in a sound body"*
- S: (adj) **good**, salutary (tending to promote physical well-being; beneficial to health) *"beneficial effects of a balanced diet"; "a good night's sleep"; "the salutary influence of pure air"*
- S: (adj) **good**, honest (not forged) *"a good dollar bill"*
- S: (adj) **good**, unspoiled, unspoilt (not left to spoil) *"the meat is still good"*
- S: (adj) **good** (generally admired) *"good taste"*

**Adverb**

- S: (adv) well, **good** ((often used as a combining form) in a good or proper or satisfactory manner or to a high standard (`good' is a nonstandard dialectal variant for `well')) *"the children behaved well"; "a task well done"; "the party went well"; "he slept well"; "a well-argued thesis"; "a well-seasoned dish"; "a well-planned party"; "the baby can walk pretty good"*
- S: (adv) thoroughly, soundly, **good** (completely and absolutely (`good' is sometimes used informally for `thoroughly')) *"he was soundly defeated"; "we beat him good"*

## 3. ALGORITHM

We estimate the semantic relatedness of two nouns distance (A, B) as follows:

• If either A or B is not a WordNet noun, the distance is infinity.
• Otherwise, the distance is the minimum length of any ancestral path between any synset v of A and any synset w of B.

In this context, we consider the word interchangeably as both words and particular two-word phrases. We proceed in the following manner.
Randomly select a document from $c_G$, the corpus of documents in the good set already classified. Call it $d_G = \{w_{1G},...,w_{nG}\}$. Similarly define $d_B = \{w_{1B},...,w_{nB}\}$.

We call $m_G^0 = \{w_{1G},...,w_{nG}\}$ and $m_B^0 = \{w_{1B},...,w_{nB}\}$.

Now pick a document, say $d_{G1}$. Calculate the semantic distance from each word in $d_{G1}$ from $d_G$ and $d_B$. If the word $w_i$ is classified to either of these groups, disregard. Else, we append the word $w_i$ to our set of good words $m_G^0$. Similarly for a "bad" document, we append words to $m_B^0$.

Proceed to cover all documents in the training set. Now, this set may have multiple redundancies. We can repeat the procedure by selecting a different choice of initial starting document $d_G$ and $d_B$. Let us thus obtain by m repetitions, the set $\{m_{Gi}\}$, $\{m_{Bi}\}$, i=1,....,m. From here we select $m_G$ to contain the k words from $\{m_{Gi}\}$ that are repeated the maximum number of times. Similarly for $m_B$.

We start with two categories of good and bad datasets, and an original collection of words for each set. We want to build up a set of words to extend our initial sets for further classification. We want to empirically determine the threshold values for each of the classes. By our algorithm, for each word to be appended to the good or the bad set, we also need to find the threshold values empirically. So, we randomly select 25000 synsets and apply our algorithm there so as to ascertain the appropriate levels of the threshold.

### 3.1 Method:

### 3.2 Detailed steps: Algorithm Part 1

Given: Two categories of dataset - A (good) and B (bad).

Category A (good): {good, dazzling, brilliant, phenomenal, excellent, fantastic, gripping, mesmerizing, riveting, spectacular, cool, awesome, thrilling, badass, moving, exciting, love, wonderful, best, great, superb, still, beautiful}

Category B (bad): {suck, terrible, awful, unwatchable, hideous, bad, clichéd, sucks, boring, stupid, slow, worst, waste}

Step 1   all:= list of all synsets
         maxp:= 0
         maxn:= 0
         pc:= 1
         nc:= 1
Step 2   i:= pc thsynset of the list all
Step 3   pc:= pc + 1
         c:= 1
Step 4   p:=cth word from the list good
Step 5   c:= c + 1
         c2:= 1
step 6:  psynsets:= list of all synsets of the word p
step 7   j:= c2th synset from the list psynsets
step 8   c2:= c2 + 1
step 9   f:= path similarity between i and j
step 10  if maxp<f then maxp:= f
step 11  if all the synsets are extracted from the list psynsets then goto step 12 else goto step 7
step 12  if all the words are extracted from the list good then goto step 13 else goto step 4
step 13  c:= 1
step 14  n:= cth word from the list bad
step 15  c:= c+ 1
         c2:=1
step 16  nsynsets:= list of all synsets of the word n
step 17  j:= c2th synset from the list nsynsets
step 18  c2:= c2 + 1
step 19  f:= path similarity between i and j
step 20  if maxn<f then maxn:= f
step 21  if all the synsets are extracted from the list nsynsets then goto step 22 else go to step 17
step 22  if all the words from the list bad are extracted then goto step 23 else goto step 14
step 23  if 0<maxp<0.8 and 0<maxn<0.2 then :
         ifmaxp>maxn then append the word part of the synseti in the list good
         else
         append the word part of the synseti in the list bad
step 24  if all the synsets of the list 'all' are extracted then goto step 25 else goto step 2
step 25  End

### 3.2.1 OUTPUT with the Final Choice of Parameter Values

['good', 'dazzling', 'brilliant', 'phenomenal', 'excellent', 'fantastic', 'gripping', 'mesmerizing', 'riveting', 'spectacular', 'cool', 'awesome', 'thrilling', 'badass', 'moving', 'exciting', 'love', 'wonderful', 'best', 'great', 'superb', 'still', 'beautiful', 'fibrillate', 'entrance', 'cathect', 'crick', 'inoculate', 'spawn', 'spat', 'infuse', 'plug', 'plug', 'seed', 'inset', 'glass', 'catheterize', 'cup', 'intersperse', 'interleave', 'feed', 'slip', 'foist', 'edit', 'tumble', 'marinade', 'decoct', 'regularize', 'tidy', 'make', 'order', 'order', 'straighten', 'rearrange', 'recode', 'reshuffle', 'serialize', 'alphabetize', 'appreciate', 'revalue', 'draw', 'arborize', 'twig', 'bifurcate',

'trifurcate', 'kill', 'unitize', 'invert', 'structure', 'restructure', 'organize', 'interlock', 'even', 'wash_down', 'synchronize', 'gauge', 'systematize', 'digest', 'codify', 'glorify', 'quantify', 'interstratify', 'stratify', 'demystify', 'ritualize', 'do_justice', 'expect', 'understand', 'extrapolate', 'involve', 'consume', 'swallow', 'take_up', 'train', 'retrain', 'drill', 'housebreak', 'toilet-train', 'indoctrinate', 'revolutionize', 'brainwash', 'drill', 'hammer_in', 'din', 'receive', 'slight', 'clear', 'misread', 'anagram', 'reread', 'dip_into', 'decipher', 'read', 'scry', 'skim', 'lipread', 'reconstruct', 'etymologize', 'quantize', 'extract', 'process', 'prorate', 'miscalculate', 'recalculate', 'average', 'square', 'cube', 'factor', 'factor', 'add', 'foot', 'subtract', 'carry_back', 'multiply', 'raise', 'divide', 'halve', 'quarter', 'differentiate', 'integrate', 'survey', 'triangulate', 'compare', 'reconsider']

..........

['suck', 'terrible', 'awful', 'unwatchable', 'hideous', 'bad', 'cliched', 'sucks', 'boring', 'stupid', 'slow', 'worst', 'waste', 'hold', 'pant', 'hack', 'palpebrate', 'wink', 'wink', 'desquamate', 'sleep', 'anesthetize', 'perk_up', 'faint', 'dimple', 'break_down', 'drop_like_flies', 'sneer', 'clear_the_throat', 'shower', 'foment', 'razor', 'marcel', 'condition', 'talc', 'bonnet', 'peel_off', 'nick', 'vegetate', 'pullulate', 'twin', 'drop', 'foal', 'alter', 'distill', 'rack', 'martyr', 'tire', 'gag', 'gnash', 'ligate', 'catch', 'catch_cold', 'hamstring', 'draw', 'suppurate', 'limber', 'give', 'give', 'follow', 'go_by', 'run_up', 'detribalize', 'change', 'rectify', 'utilize', 'gentrify', 'republish', 'defoliate', 'disbud', 'freeze-dry', 'tin', 'slack', 'air-slake', 'flow', 'lave', 'reduce', 'tie', 'gate', 'draw_the_line', 'consolidate', 'reflate', 'paralyze', 'freeze', 'ablate', 'predate', 'peroxide', 'ebonize', 'habilitate', 'rescale', 'pan-broil', 'stave_in', 'obstinate', 'expatriate', 'rush', 'morph', 'cancel_out', 'accommodate', 'harmonize', 'glue', 'water_down', 'gauge', 'scale', 'meter', 'isolate', 'reline', 'lather', 'seethe', 'cut_in', 'sentimentalize', 'superannuate', 'taste', 'relive', 'understudy', 'mistake', 'proofread', 'put_out_feelers', 'plumb', 'prospect', 'google']

### 3.3 ALGORITH Part 2

In this part we calculate the frequency of the list of words that we have obtained from the expansion in the previous section. The following is a detailed algorithm for the process.

### 3.3.1 Algorithm

Given:  i) a list named 'keywrds' containing a list of words, used to create a set of sublinks

ii) a list named 'searchwrd' containing a list of words whose frequency will be calculated

Step 1   extract all the links present in the content of a given url and store them in a list 'links'

Step 2   set i: =1

Step 3   extract ith link from the list 'links'

Step 4   set j: =1 and i: =i+1

Step 5   extracts jth keyword from the list 'keywrds'

Step 6   if the keyword is a substring of the link (i.e. if the link contains the keyword ) then the link is appended into a list called 'sublinks' and then goes to step 9

Step 7   set j: = j + 1

Step 8   if all the keywords are extracted from the list 'keywrds' then go to step 9 else go to step 5

Step 9   if all the links are extracted from the list 'links' then go to step 10 else go to step 3

Step 10  initialize all the elements of a 2D matrix 'freq [][]' by value 0

Step 11  set i: =1

Step 12  extract the ith link from the list 'sublinks'

Step 13  the content of the ith link is stored in a variable named 'data'

Step 14  set j: = 1

Step 15  extract jth line from 'data' and store the line in a variable named 'line'

Step 16  set j: = j + 1

Step 17  set k:=1 and l: = 1

Step 18  extract the lth word from the given list 'searchwrd'

Step 19  count the number of times the word occurs in that line and store the count in variable 'p'

Step 20  compute $freq_{i,k} = freq_{i,k} + p$

Step 21  set k: = k + 1 and l: =l + 1

Step 22  if all the words from the list 'searchwrd' are extracted then go to step 23 else go to step 18

Step 23  if all the lines from 'data' are extracted then go to step 24 else go to step 15

Step 24  if all the links from the list 'sublinks' are extracted then go to step 25 else go to step 12

Step 25  end

## 4. DISCUSSION

We have used the function wn.path_similarity(synset1,synset2)which returns a score denoting how similar two word senses are, based on the shortest path that connects the senses in the is-a (hypernym/hypnoym) taxonomy. The score is in the range 0 to 1. A score of 1 represents identity i.e. comparing a sense with itself will return 1.

Example, wn.path_similarity(hit, slap)returns 0.142

Here we have taken a total of 25000 words from the WordNet. Different choices for the tau value in the condition check statement was tried for the balance and usability of the output classifier. After several trials with value pairs (0.5,0.5), (0.7,0.3), (0.7,0.5), (0.8,0.1), (0.8,0.3), (0.8,0.5), it was finally empirically adjusted to 0.8 and 0.2 for good and bad sets of word respectively so that the final two sets contain approximately same number of words. Now these words are considered as a "basis" for the two sets and will be used to classify any content based document. We need to have the "good" and "bad" sets of approximately similar sizes so that classification in either set is not more rigorous than in the other. If, for example, we have the "good" set with a cardinality much higher than that of the "bad" set, we will find it easier to classify to "good" sets than to "bad" ones.

We will apply our classification method to a corpus with small size of documents where bag of words is expected to not work well.

## 5. First Level Analysis

We apply our mechanism on Wikipedia entries on books published on 1970 to understand what sort of reviews the books garnered. However, only 12 of the books have fields titled "Reception", and hence only those can be analysed. On frequency calculation for the good and bad sets of the words, we notice that the output vectors are highly sparse, indicating that the authors of the said documents have a somewhat biased vocabulary while describing similar sentiments.

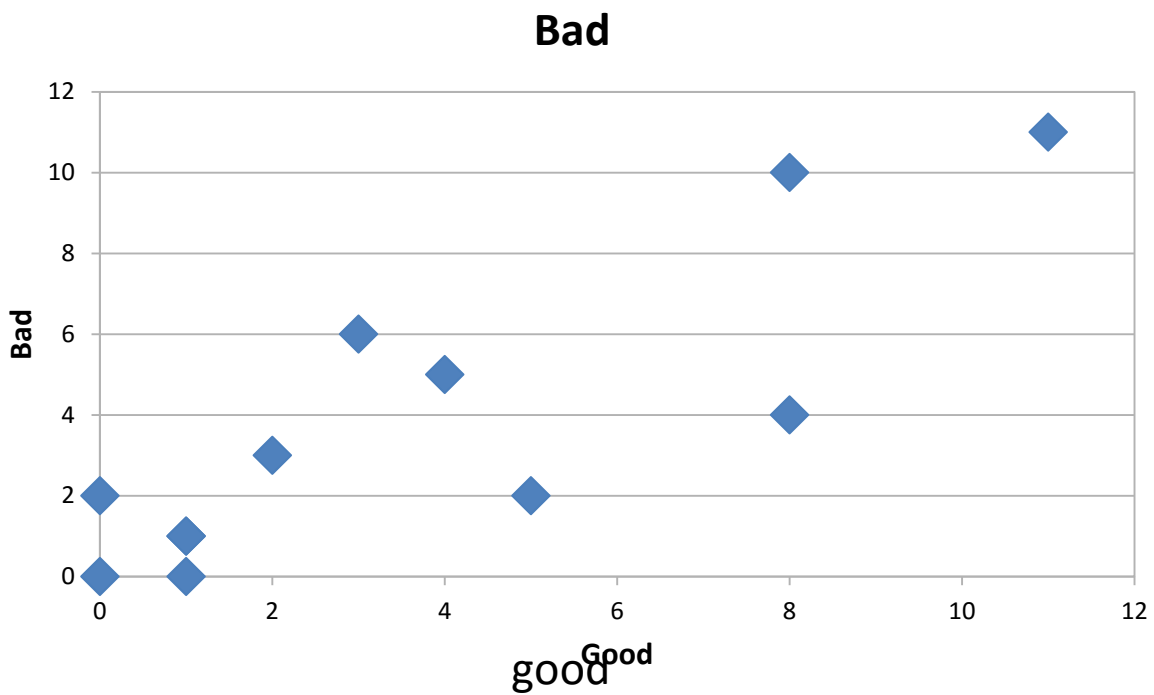

Figure 3: Graphical Representation of Results

The above graph represents the final results. For each point in the X-axis, the value represents the frequency obtained for the good set and the Y-axis represents the frequencies for bad sets.

Ideally we would like to obtain a clear separation in the plane for the data between the good and the bad sets for easy classification.

**5.1 Further Guard against Misclassification:**

We have been disregarding the order of words in this problem. However, this may lead to misconceptions about the sentiment conveyed in a sentence. For example, the presence of a conjunction like "but" or "however" may indicate a reversal of sentiment from the first part to the second part of a sentence, while "and" and "also" may lend more credence to the sentiment. Thus, via these words, any compound sentence can be broken down into a sequence of {+1,-1} where +1 indicates sentiment towards a good set and -1 that towards a bad. An average of these signs may give us more idea about the appropriate classification.

**5.2 Further Works**

By cross-validation, an initial choice of good and bad words may be chosen from pre-classified documents. Those sets can be expanded via the illustrated method and then used for classification.

**7. Appendix:**

**7.1 PROGRAM for the Algorithm Part 1**

```python
from nltk.corpus import wordnet as wn
def func(m):
    st=m.name()
    st=str(st)
    print st
    i=0
    wrd=""
    l=len(st)
    while i<l:
        ch=st[i]
        if ch!='.':
            wrd=wrd+ch
            i=i+1
        else:
            print wrd
            return wrd

all=list(wn.all_synsets())
max=0
maxn=0
c1=0
c=0
c2=0
pos=['good','dazzling','brilliant','phenomenal','excellent','fantastic','gripping','mesmerizing','riveting','spectacular','cool','awesome','thrilling','badass','moving','exciting','love','wonderful','best','great','superb','still','beautiful']
neg=['suck','terrible','awful','unwatchable','hideous','bad','cliched','sucks','boring','stupid','slow','worst','waste']
for i in all:
        c=c+1
        c1=0
        ct=0
        for p in pos:
                c1=c1+1
            c2=0
            psynsets=wn.synsets(p)
            for j in psynsets:
                c2=c2+1
                f=wn.path_similarity(i,j)
                if(max<f):
                    max=f
                if(c2==3):
                    break

        for n in neg:
            c2=0
            nsynsets=wn.synsets(n)
            for j in nsynsets:
                c2=c2+1
                f=wn.path_similarity(i,j)
                if(maxn<f):
                    maxn=f
                if(c2==5):
                    break
```

```
        if max<0.8 and max>0 and maxn<0.2 and maxn>0:
            w=func(i)
            if max>maxn:
                pos.append(w)
            else:
                neg.append(w)

        max=0
        maxn=0
        if c==25000:
            break

print pos
print ".........."
print neg
```